\documentclass[
 reprint,
 amsmath,amssymb,
 aps
]{revtex4-2}

\usepackage{graphicx}
\usepackage{dcolumn}
\usepackage{bm}
\usepackage[latin1]{inputenc}
\usepackage{upgreek}

\begin{document}

\preprint{APS/123-QED}

\title{Locomotion of magnetoelastic membranes in viscous fluids}

\author{Chase Austyn Brisbois}
 \affiliation{Department of Materials Science and Engineering, Northwestern University, Evanston, IL 60208}
 
\author{M\'{o}nica Olvera de la Cruz}
 \email{m-olvera@northwestern.edu}
 \affiliation{Department of Materials Science and Engineering, Northwestern University, Evanston, IL 60208.}
 \affiliation{Department of Chemistry, Northwestern University, Evanston, IL 60208.}
 \affiliation{Department of Physics and Astronomy, Northwestern University, Evanston, IL 60208.}

\date{\today}

\begin{abstract}
The development of multifunctional and biocompatible microrobots for biomedical applications relies on achieving locomotion through viscous fluids. Here, we describe a framework for swimming in homogeneous magnetoelastic membranes composed of superparamagnetic particles. By solving the equations of motion, we find the dynamical modes of circular membranes in precessing magnetic fields, which are found to actuate in or out of synchronization with a magnetic field precessing above or below a critical precession frequency, $\omega_c$, respectively. For frequencies larger than $\omega_c$, synchronized rotational and radial waves propagate on the membrane. These waves give rise to locomotion in an incompressible fluid at low Reynolds number using the lattice Boltzmann approach. Non-reciprocal motion resulting in swimming is achieved by breaking the morphological symmetry of the membrane, attained via truncation of a circular segment. The membrane translation can be adapted to a predetermined path by programming the external magnetic field. Our results lay the foundation for achieving directed motion in thin, homogeneous magnetoelastic membranes with a diverse array of geometries. 
\end{abstract}

\maketitle

\section{INTRODUCTION}
\begin{figure}
\centering
\includegraphics[scale=0.5]{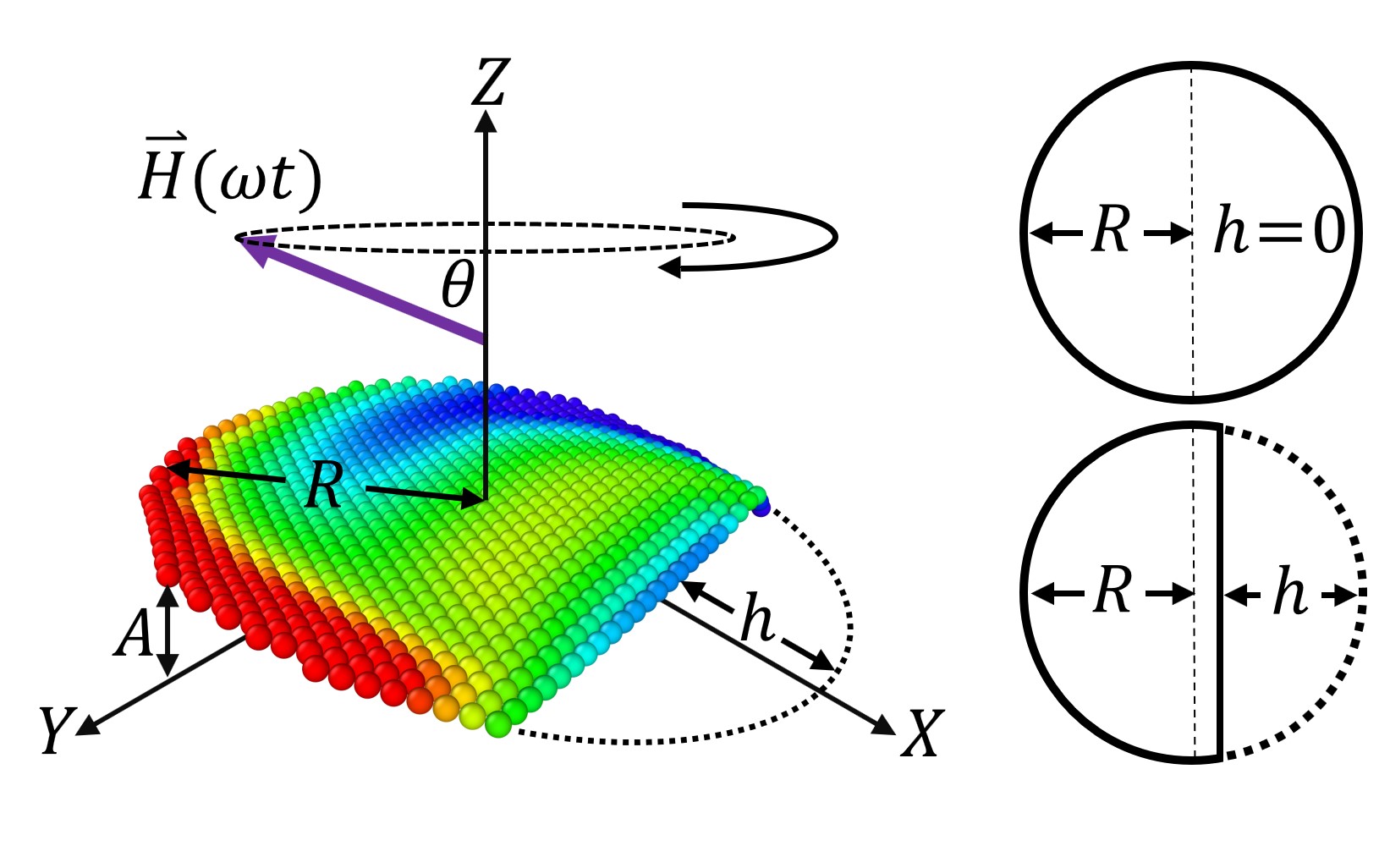}
\caption{An image of a truncated magnetoelastic membrane in a precessing magnetic field. The degree of truncation $S=h/2R$, where $h$ is the sagitta length of the removed circular segment, and $R$ is the membrane radius, determines membrane symmetry. The magnetic field $\vec{\bm{H}}$ precesses at the angle $\theta$ around the $z$-axis with a phase given by $\phi=\omega t$, where $\omega$ is the precession frequency and $t$ is time. The field induces an amplitude, $A$, along the membrane perimeter, measured from the $x$-$y$ plane. Coloration indicates z-position (+z in red and -z in blue).}
\label{fig:fig1}
\end{figure}

\begin{figure*}
\centering
\includegraphics[scale=0.62]{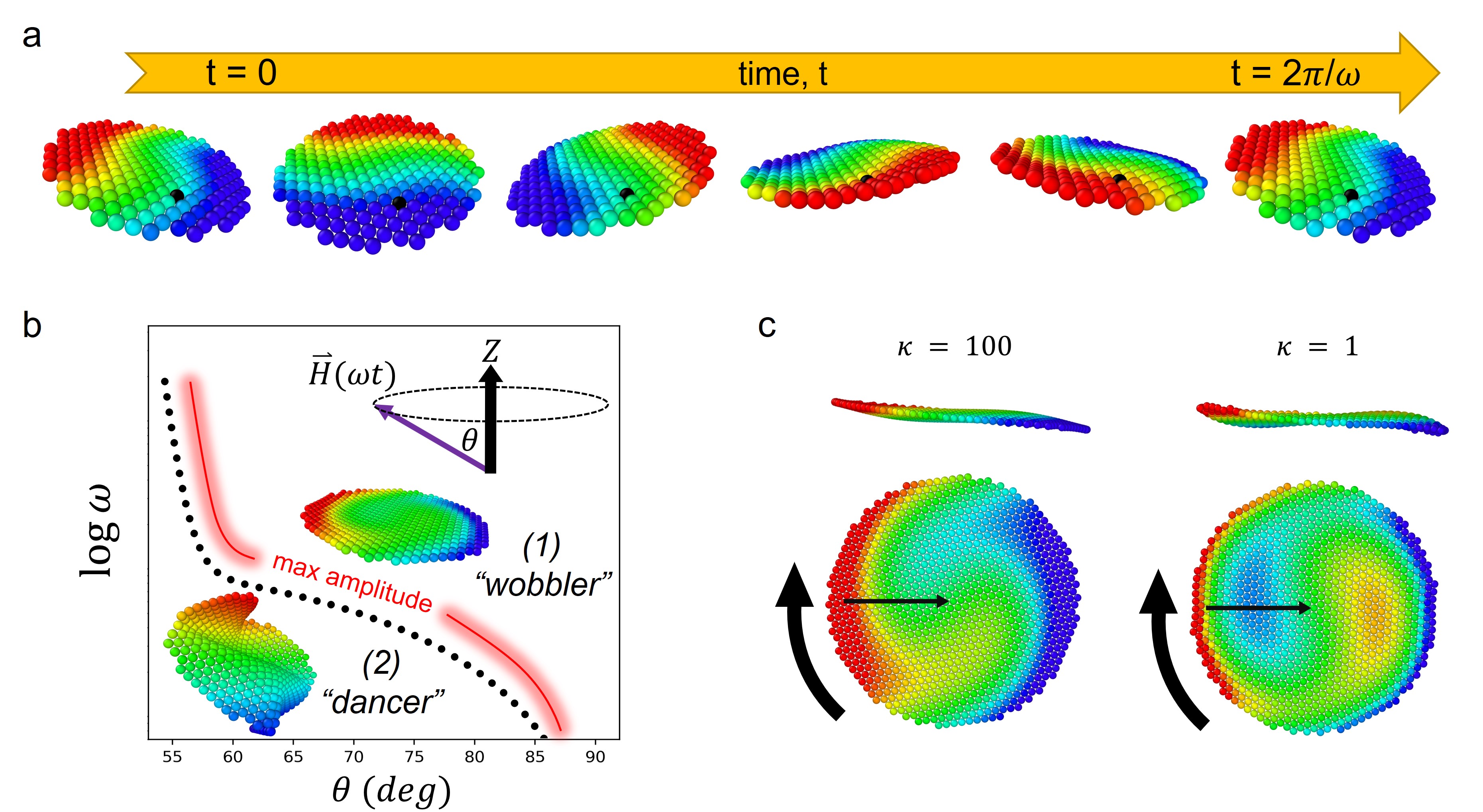}
\caption{A circular magnetoelastic membrane in a precessing magnetic field adopts dynamic motion. (a) Transverse waves propagate around the membrane above a critical frequency ($\omega > \omega_c$). (b) A schematic plot showing the phase diagram of a membrane. Above the dotted black curve, the ``wobbling" membrane remains perpendicular to the precession axis and possess the rotational waves from (a). The wave amplitude maximizes just before the transition. Below this curve, the membrane buckles and rotates asynchronously with the field, hence ``dancers". (c) The bending stiffness controls the shape of the rotational waves. The black arrows indicate the direction of wave propagation. Coloration indicates z-position (+z in red and -z in blue).}
\label{fig:fig2}
\end{figure*}

Magnetically controlled microrobots have applications in drug delivery  \cite{klosta2020kinematics,Dreyfus2005micro, jang2015undulatory, bryan2019magneto}, sensing  \cite{moerland2019rotating,goubault2003flexible}, micromixing  \cite{biswal2004mixing}, detoxification  \cite{zhu2015microfish, wu2015nanomotor} and microsurgery  \cite{wu2020multifunctional, Vyskocil2020cancer}. Such versatile use of magnetic microrobots is possible because magnetic fields can penetrate organic matter, do not interfere with biological or chemical functions, do not require fuel, and, most importantly, can be externally controlled. These properties allow for non-invasive and precise spatiotemporal execution of desired function. In particular, superparamagnetic particles are ideal candidates for robotic functions when combined with elastic components \cite{dempster2017contractile,Dreyfus2005micro} due to their lack of residual magnetization, lowering their propensity to agglomerate, and are less toxic than ferromagnetic particles \cite{markides2012biocomp}. Magnetoelastic membranes possess a diverse repertoire of possible dynamic motion under time-dependent magnetic fields  \cite{brisbois2019actuation,Hu2018small}, making them particularly suited for designing multifunctional microrobots.

Navigating a viscous environment requires a magnetoelastic microrobot to use competing magnetic and elastic interactions to induce non-reciprocal motion  \cite{purcell1977life}. That is, the sequence of configurations that the robot adopts must break time-reversal symmetry to swim at low Reynolds numbers (Re $\ll 1$). For example, magnetoelastic filaments achieve non-reciprocal motion with a non-homogeneous distribution of magnetic components or with shape asymmetry  \cite{Dreyfus2005micro, jang2015undulatory, bryan2019magneto, yang2020reconfig, cebers2005flexible}. These asymmetries induce bending waves that propagate along the chain. 

In nature, microscopic organisms such as euglenids  \cite{Arroyo2012reverse} swim using self-propagating waves directed along their cellular membrane. G. I. Taylor was the first to model such organisms using a transverse wave traveling along a infinite 2D sheet  \cite{taylor1951analysis}. Taylor found that the wave induced motion in the sheet opposite to the propagating wave direction. Subsequent works expanded upon Taylor's findings  \cite{lauga2009hydro}, and developed a rotational counterpart  \cite{Corsi2019neutrally} that produces a hydrodynamic torque on circular membranes with prescribed waves traveling around their perimeter.

In this article, we study rotational waves in homogeneous superparamagnetic membranes under precessing magnetic fields. We investigate the dynamic modes of the membrane separated by a critical precession frequency, $\omega_c$, below which the membrane motion is asynchronous with the field, and above which rotational waves propagate in-phase with the field precession. Breaking the membrane's center of inversion symmetry, by removing part of the circle (Fig 1), allows for locomotion in the fast frequency phase ($\omega > \omega_c$). Shape asymmetry is needed to disrupt the inversion symmetry of the magnetic forces experienced by a circular membrane.
We show that the torque and velocity of the membrane counterintuitively resembles the linear Taylor sheet rather than its rotational analogue. 
Furthermore, by controlling a magnetoviscous parameter and the membrane shape asymmetry, we demonstrate swimming directed by a programmed magnetic field and diagram its non-reciprocal path through conformation space. 

The paper is organized as follows. In Sec. II, we establish the phase diagram of a circular magnetoelatic membrane in a precessing magnetic field and determine the transition frequency $\omega_c$. In Sec. III, we introduce hydrodynamic interactions and observe circular locomotion in asymmetric membranes. We demonstrate a programmed magnetic field, in Sec. IV, that directs a membrane swimmer along a predetermined path. Finally, we make concluding remarks on the necessary conditions for superparamagnetic swimmers in Sec. V.

\section{PHASE SPACE FOR UNTRUNCATED MEMBRANE}

\begin{figure}
\centering
\includegraphics[scale=0.65]{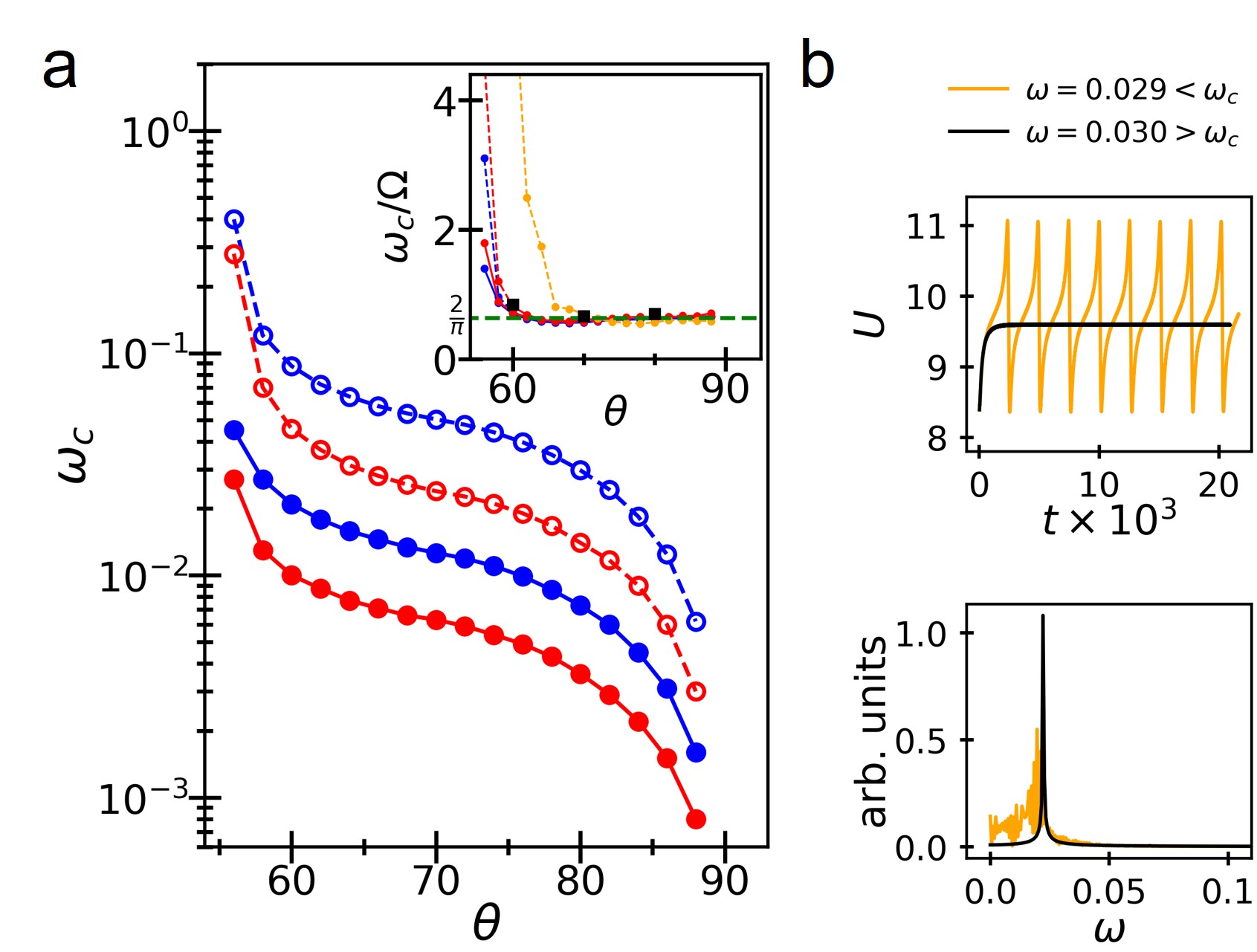}
\caption{The synchronous-asynchronous (wobbler-dancer) transition frequency $\omega_c$ for a magnetoelastic membrane. (a) Molecular dynamics calculation of $\omega_c$ as a function of the field precession angle $\theta$. The solid and dashed lines indicate a dipole magnitude of $\mu=$ 2 and $\mu=$ 1, respectively. The inset shows the dimensionless transition frequency $\omega_c/\Omega$, where $\Omega$ is the membrane's characteristic rotation frequency. The green dashed line represents the theoretical transition at $\omega_c/\Omega=2/\pi$, which, near $\theta=$ 90$^\circ$, is independent of bending stiffness ($\kappa=$ 1, orange. $\kappa=$ 100, blue/red). The black squares show the transition calculated from lattice-Boltzmann simulations. (b) Supercritical and subcritical behavior of the total energy U (magnetic + bending). The precession frequency is close to the critical frequency, $0.029 < \omega_c < 0.030$ ($\theta=$ 80$^\circ$). Fourier transform of the rotational wave amplitude (bottom).}
\label{fig:fig3}
\end{figure}

We construct the phase diagram for the dynamic modes of the membrane using molecular dynamics (MD) without hydrodynamics to efficiently search for non-reciprocal actuation relevant to locomotion. Actuation of magnetoelastic membranes in time-dependent magnetic fields necessitates a model that captures elastic bending in response to magnetic forces, which are imparted by the dipolar interactions of embedded magnetic colloids. The membrane is composed of a hexagonal close-packed monolayer of hard spherical colloids, each of diameter $\sigma$ and possessing a point dipole moment $\bm{\mu}$ at its center. The bonds between the colloids are approximately inextensible, but able to bend with rigidity, $\kappa$. We model an implicit, uniform magnetic field by constraining the orientation of the colloids' dipole moments in the direction of the field, $\bm{H}=\bm{\mu}/\chi$, where $\chi$ is the magnetic susceptibility of the material and $\bm{\mu}$ is the dipole moment with magnitude $\mu$. The instantaneous dipole orientation is given by $\bm{\hat{\mu}} = \sin{\theta}\sin{\omega t}~\bm{\hat{i}} + \sin{\theta}\cos{\omega t}~\bm{\hat{j}} + \cos{\theta}~\bm{\hat{k}}$, where $\theta$ is the field precession angle, $\omega$ is the precession frequency, and $t$ is time. All quantities herein are expressed in dimensionless units (see Appendix A).

A diverse set of possible dynamic motion develop depending on the radius $R$ of the thin membrane and the magnetic field parameters ($\mu$, $\theta$, $\omega$). While varying these parameters, we solve the equations of motion for an overdamped system with a friction force imparted on each colloid given by $-\xi v(t)$, where $v(t)$ is the colloid velocity, and $\xi$ is the damping coefficient. Within this approximation, two dynamic mode regimes develop. At fast frequencies ($\omega > \omega_c$), the membrane motion synchronizes with the field to produce transverse waves that propagate around the membrane (Fig. 2a). At slow frequencies ($\omega < \omega_c$), we observe a collection of modes that are asynchronous with the field. We find a critical frequency, $\omega_c$, where there is an abrupt change in the membrane's dynamic motion (Fig 2b).

As the field precesses, the forces along the membrane perimeter generate internal buckling and create a torque that rotates the membrane around its diameter. If the magnetic field precession is fast ($\omega > \omega_c$), the continuous change in the direction of the axis of rotation leads to the development of a constant-amplitude wave traveling along the membrane perimeter, see Video 1 in Ref. \cite{video1}. On average, the membrane remains perpendicular to the precession axis and simply ``wobbles", synchronous to the field, and with no significant rotation around the precession axis. This state closely resembles acoustically levitated granular rafts \cite{lim2021acoustically}. 

The direction of the propagating wave matches the handedness of precession because the dipole-dipole forces, which cause buckling, point in the direction of the magnetic field. However, the field polarity does not affect the magnitude or travel direction of the wave since the superparamagnetic dipoles are always oriented in the same direction as the field. Hence, the force due to the dipole-dipole interactions, ${\bm{F}}_{dipole}$, remains unchanged (${\bm{F}}_{dipole} \propto (\bm{\mu} \cdot \bm{r})\bm{\mu} = (-\bm{\mu} \cdot \bm{r})(-\bm{\mu})$, where $\bm{r}$ is the displacement vector between dipoles  \cite{yung1998analytical}).

In addition to the rotational waves, the wobbling mode also manifests radially propagating (inward) bending waves (Fig. 2c) that terminate at the membrane center. The wave shape weakly depends on the membrane stiffness $\kappa$; the wave form is better defined as $\kappa$ decreases. However, totally compliant membranes ($\kappa\rightarrow 0$) do not transmit bending waves and therefore this phenomenon exists only for intermediate $\kappa$.	
\begin{figure*}
\centering
\includegraphics[scale=0.65]{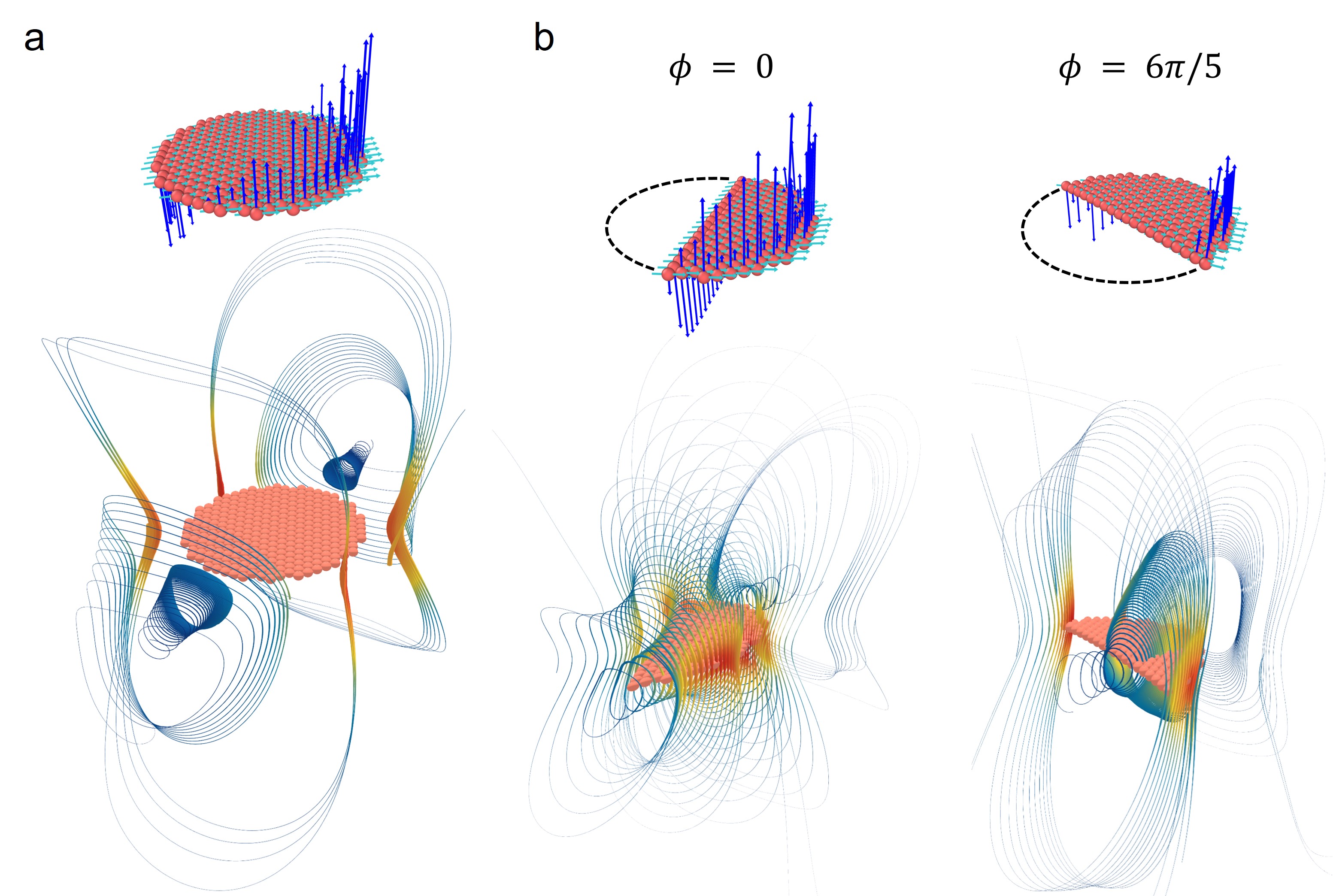}
\caption{Fluid flow around a magnetoelastic membranes in the ``wobbler" regime. The top images show the total force vector for each colloid (blue arrows) alongside the dipole orientation (cyan arrows) for a precessing field ($\mu=$1, $\theta=$ 70$^\circ$, $\omega=$ 0.1). The bottom images show streamlines around the membrane, where the color indicates flow speed (red $>$ blue). (a) A snapshot of a circular membrane. (b) Two snapshots of a truncated circular membrane separated by a shift in the field precession $\Delta\phi=\omega t=6\pi/5$.}
\label{fig:fig4}
\end{figure*}

If the precession is slow ($\omega < \omega_c$), the membrane has enough time to rotate completely parallel to the precession axis and will adopt new configurations due to elastic buckling. How the membrane buckles depends on the magnetoelastic parameter  \cite{vazquez2018flexible}, $\Gamma = M L^2 / \kappa$, which characterizes the ratio between the membrane's magnetic and bending energies, where $M$ is the magnetic modulus, and $L^2$ is the membrane area. If the magnitude of $\Gamma$ is very small ($\Gamma \ll 1$) or very large ($\Gamma \gg 1$), we observe hard disk behavior because bending distortions become impossible due to mechanical stiffness or due to unfavorable magnetic interactions, respectively. While not investigated here, strong magnetic coupling \cite{park2020dna,messina2015quant} between colloids will adversely affect membrane synthesis. 

At intermediate $\Gamma$, membrane edges buckle several times per precession period and produce magnetically stabilized conformations that, while periodic, run out-of-sync with the field, see Video 2 in Ref. \cite{video2}. Much of this back-and-forth ``dancing" motion is essentially reciprocal and is therefore generally a poor candidate for studying swimming at small Re. Therefore, here we seek to formally define $\omega_c$ and focus on the wobbling regime ($\omega > \omega_c$).

To accurately determine the transition frequency $\omega_c$ that separates the wobblers from the dancers, we investigate how the magnetic field parameters (precession angle $\theta$, dipole magnitude $\mu$), and membrane radius $R$ (Fig. 3a) contribute to the characteristic response time $\tau$ of the rotating membrane. When the membrane rotation time $\tau$ increases, it necessarily requires a slower field to keep the membrane in the wobbling mode, decreasing $\omega_c$. A larger $\tau$, can be achieved by weakening the magnetic torque ($\theta$ closer to $\pi/2$ or smaller $\mu$) or increasing the drag on the membrane (larger $R$). Similarly, a smaller $\tau$ implies a fast membrane response from a strong field or a small membrane. We observe that $\omega_c$ diverges as $\theta$ approaches the magic angle, partly due to instability of the wobbling phase at angles below the magic angle  \cite{cimurs2013dynamics}.

The transition to the wobbling state is characterized by the abrupt shift in the membrane's potential energy from a time-dependent function to a constant value (Fig. 3b, top). When the potential energy does not change, this implies that the shape of the membrane conformation becomes invariant in the rotating field reference frame. This change in the dynamic buckling results in a single Fourier mode for the displacement of the colloids parallel to the precession axis (Fig. 3b, bottom). This resembles the transition between the synchronous and asynchronous motion for oblate magnetic particles  \cite{cimurs2013dynamics}.

\begin{figure*}
\centering
\includegraphics[scale=0.65]{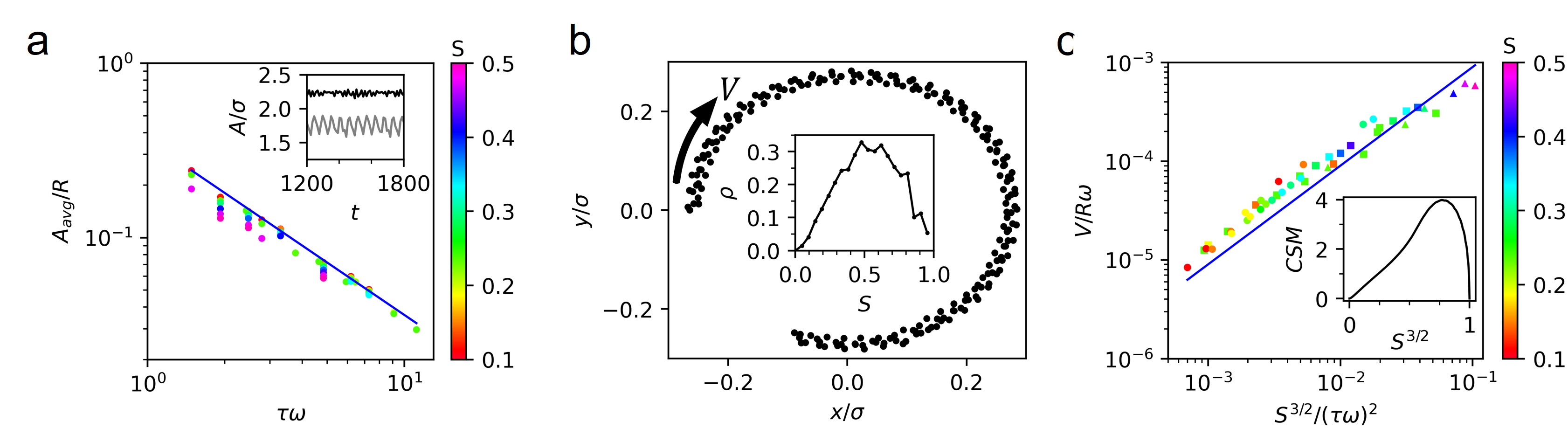}
\caption{Actuation drives circular locomotion of truncated magnetoelastic membranes through a viscous fluid. (a) The average rotational (``wobble") wave amplitude $A_{avg}$, scaled by the membrane radius $R$, depends inversely on the magnetoviscous parameter $\tau \omega$. Data points from lattice Boltzmann simulations are compared to our analytical model (solid blue line). The coloration of the simulation data notes the degree of truncation $S$. The inset shows the variation in $A/\sigma$ over time based on membrane geometry ($S=$ 0.05, black; $S=$ 0.5, gray), where $\sigma$ is the colloid diameter. (b) The path taken by a membrane in a precessing field. The arrow indicates the travel direction with velocity $V$. The inset shows the radius $\rho$ of this path as a function of $S$. (c) The membrane velocity is proportional to $A_{avg}^2 \propto (\tau \omega)^{-2}$ and scales with $S^{3/2}$ due to changes in the length of the membrane perimeter. The data points shape are coded by the membrane radius ($R=$ 7, triangle; $R=$ 9, square; $R=$ 12, circle). The line shows our analytical prediction (slope = 1.0). The inset shows the continuous inversion symmetry measure for a flat truncated membrane.}
\label{fig:fig5}
\end{figure*}

When the precession angle approaches $\pi/2$, the membrane motion becomes independent of the stiffness of the membrane; the membrane remains flat at all times and for all values of $\omega$. As the field precesses, the forces perpendicular to the membrane plane vanish near $\theta = \pi/2$ preventing significant radial bending and, consequently, changing $\kappa$ does not shift $\omega_c$ (Fig. 3a, inset). By solving an Euler-Lagrange equation with Rayleigh dissipation (Appendix B), we derive an equation of motion for a membrane in a field precessing at a large angle. It reveals a characteristic frequency of membrane motion, $\Omega = 6 \zeta(3)\mu_0\mu^2\sin{2\theta} / \pi^2 \eta R^2 \sigma^4$, where $\mu_0$ is the magnetic permeability of free space, $R$ is the radius of the membrane, $\eta$ is the viscosity, and $\zeta(x)$ is the Riemann zeta function. The frequency $\Omega$ comes from the magnetic ($\propto \mu_0\mu^2\sin{2\theta} R^2/\sigma^5$) and drag ($\propto \eta R^4/\sigma$) potential functions. The $\omega_c$ curves in Fig. 3a can be scaled by $\Omega$ to obtain a dimensionless transition frequency $\omega_c / \Omega = 2/\pi$ (Fig 3a, inset). This provides a single number with which to predict the dynamic motion of a membrane and defines the membrane response time $\tau = \Omega^{-1}$.

\section{HYDRODYNAMIC EFFECTS ON ``WOBBLING" MEMBRANES}

It is useful to investigate the broad range of dynamic motions accessible to a magnetoelastic membrane using a simple overdamped system to highlight relevant transitions in motion. Afterwards, we confirmed the dimensionless transition $\omega_c / \Omega$ using the more computationally expensive hydrodynamic simulations (Fig. 3a, black squares) and change the characteristic frequency $\Omega = 27 \zeta(3)\mu_0 \mu^2\sin{2\theta}/64\eta R\sigma^5$ to include hydrodynamic interactions (see Appendix B). Using the same magnetic potential as the previous section, this change in $\Omega$ is due to the torque on the membrane a viscous fluid ($\propto \eta R^3$). We will use this definition for $\Omega$ hereafter.

To observe the effect of the wobbler's non-reciprocal motion on the surrounding fluid, we add hydrodynamic interactions to our simulations by coupling the MD model to the lattice Boltzmann method (LBM)  \cite{Mackay2013hydrodynamic}. This technique, which comes from a discretization of the Boltzmann transport equation, reproduces the incompressible Navier-Stokes equation in the macroscopic limit. LBM calculates the evolution of a discrete-velocity distribution function, $f_i$, at each fluid node that fills the simulation box on a square lattice mesh with a spacing of $\Delta x$. The surface of the colloids act as a boundary and is defined by surface nodes that interact with the fluid using the model developed by Peskin  \cite{peskin2002immersed}. Care must be taken when implementing LBM with MD because compliant springs can cause translation of the membrane due to in-plane stretching. This mechanism has been observed in systems of a few colloids  \cite{Grosjean2018surface}. Therefore, the stiffness of the springs must be large enough to eliminate this effect for an inextensible membrane model requiring the use of a smaller simulation timestep. Simultaneously, small Re in LBM is achieved by decreasing the Mach number, set by the speed of sound $c_s=\frac{1}{\sqrt3}\frac{\Delta x}{\Delta t}$  \cite{kruger}. Therefore, we rely on a small timestep that is compatible with both schemes. See Appendix C for a complete description of the model.

The fluid flow around the membrane is determined by its symmetry and actuation. The precessing magnetic field induces a torque along the membrane perimeter that circulates fluid around an axis of rotation through the membrane diameter (Fig. 4a). This axis of rotation moves continuously with the field, producing circulating flows above and below the membrane. The rising peaks push the fluid up (+z) and it flows towards the falling peak (-z) on the other side of the membrane. This flow simultaneously resembles analytical predictions for rotating hard disks \cite{tanzosh1996general} and the flow vorticity from Taylor's swimming sheet \cite{lauga2009hydro}. 

\begin{figure*}
\includegraphics[scale=0.5]{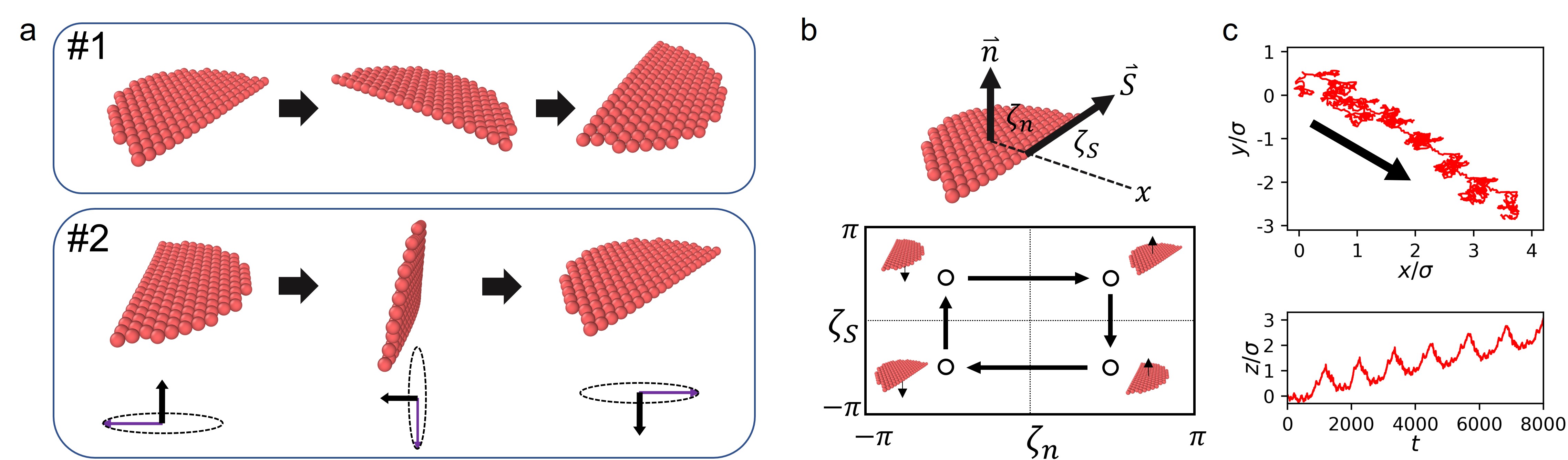}
\caption{A two-step magnetic field directs a swimming membrane along a path. (a) First, a membrane wobbler moves under a precessing magnetic field. After it rotates a half-turn (\#1), the precession switches to a fast frequency at $\theta=\pi/2$ while the axis rotates to flip the membrane (\#2). (b) We define the angles that the normal vector $\bm{n}$ and the truncation vector $\bm{S}$ make with the $x$-axis as $\zeta_n$ and $\zeta_S$, respectively. (c) The path in conformation space over the two-step field. (c) Repeated cycles from (a) move the membrane against the Brownian motion of a thermalized fluid. The upper panel shows the motion of the membrane in the $x$-$y$ plane. The black arrow indicates the direction of motion. The lower panel shows the displacement in the $z$-direction.}
\label{fig:fig6}
\end{figure*}

The centrosymmetry of a magnetoelastic membrane generates a flow that prevents its center of mass from translating. To induce locomotion, we truncate the membrane by removing a circular segment with a sagitta of length $h$. We normalize $h$ by the diameter of the circle to define the degree of truncation of the circular membranes as $S = h/2R$.  In contrast to the circular membrane case, the shape of the fluid flow in the truncated membrane changes during a single precession period leading to asymmetric flow field depending on the relative orientation between the field and the truncation cut (Fig. 4b).

The amplitude of propagating waves is particularly relevant for predicting the translational \cite{taylor1951analysis} or rotational \cite{Corsi2019neutrally} velocity of a membrane. Here, the wobble amplitude can be calculated by balancing the magnetic \cite{yung1998analytical} and drag \cite{tanzosh1996general} torque in a viscous fluid (see Appendix D). Under small amplitudes for the rotational wave, we obtain the simple relation
\begin{equation}
    \frac{A}{R}=\frac{C}{\tau\omega}
    \label{eq:amplitude}
\end{equation}
where $A$ is the amplitude, and $C=32/9\pi^2$ (Fig. 5a). It is reasonable to assume, in the limit of small deformations, the bending contribution to the torque along the edge is negligible, unless $\kappa \rightarrow \infty$. Furthermore, the amplitude is independent of the membrane size since $\tau \propto R$. However, the membrane is not free to increase in radius arbitrarily. The small Re condition implies that $\nu \gg R^2/\tau$, where $\nu$ is the kinematic viscosity. Obeying this constraint on $\tau$, we can define a magnetoviscous parameter $\tau \omega$ and use it to predict locomotion of the membrane. 

Asymmetry in the fluid flow due to a degree of truncation $S>0$ leads to locomotion of the membrane. The membrane travels with a net velocity in the direction of the truncation cut. This net motion is due to the decrease in the amplitude of the waves traveling along the truncated edge. Since the truncated edge is closer to the center of mass and $\kappa$ is homogeneous, the membrane will bend to a lesser extent along the truncation. This manifests as a net motion every $2\pi/\omega$, where reversing the handedness of the field reverses the locomotive direction. 

However, the membrane follows a curved path. While the membrane torque due to the underlying colloidal lattice is negligible, the membrane can rotate significantly by choosing $\omega$ close to $\omega_c$. This rotation emerges exclusively due to the magnetic interactions perpendicular to the wobbling membrane. If the projection of the forces, visualized in Fig. 4, on the $x$-$y$ plane are non-zero, the membrane will rotate. Over many precession periods, the membrane moves in a circular path around a central point (Fig. 5b). The radius $\rho$ of the path depends on $S$ and $A_{avg}$. Untruncated, $S=0$, and fully truncated, $S = 1$, do not translate and result in $\rho=0$. Hence, a maximum for $\rho$ exists at intermediate S values (Fig. 5b inset). Since the membrane is composed of colloids, irregularities in the $\rho \left(S,A_{avg}\right)$ curve appear because the symmetry of the membrane changes in discrete steps.

The magnetic field controls how quickly the membrane travels along the circular path and affects its angular velocity. Together with the truncation $S$, the velocity $V$ at which the membrane translates along the path can be determined using a singularity method. With a nearest-neighbors assumption for the magnetic interactions and treating them as point-disturbances, the advective flow through the center of mass leads to the velocity
\begin{equation}
    \frac{V}{R\omega} = \frac{C^2}{12 \zeta(3)} \frac{S^{3/2}}{{(\tau\omega)}^2}
    \label{eq:velocity}
\end{equation}
where the $S^{3/2}$ dependence comes from the number of uncompensated point forces formed by truncation. The velocity is normalized by the phase speed $R\omega$. The inverse squared relation on $\tau \omega$ for the velocity is a result of the dependence on the product of the magnetic force and wave amplitude, which, in turn, relies on the magnetic force. Here, we recover the velocity dependence on the square of the wave amplitude \cite{taylor1951analysis}, but with a lower velocity ($V \leq V_{Taylor}/6$). Appendix E contains the full derivation. We see a deviation from the relationship obtained in Eq. \ref{eq:velocity} at large values of $S^{3/2}/(\tau \omega)^2$ owing to either a high degree of truncation (a linear polymer) or a small viscomagnetic parameter (``dancer") (Fig. 5c). The direction of $V$ is dictated by the handedness of the precessing field and is an example of magnetically induce symmetry breaking. We find that the continuous symmetry measurement \cite{zabrodsky1992continuous} can predict relative changes in the velocity of locomotion. When the inversion asymmetry increases, $V$ increases (Fig. 5c, inset) because the conformational path taken by the membrane widens, leading to greater net work done on the fluid \cite{grosjean2016realization}.

\section{MEMBRANE SWIMMING}

Here, we give an example of how a programmed magnetic field can produce a non-reciprocal conformational path that results linear swimming. In Fig. 6a, we show that a precessing field can rotate the membrane 180$^\circ$ from its initial configuration. Then, the precession frequency is increased and $\theta$ is set to $\pi/2$. This keeps the membrane flat in the precession plane while the precession axis is rotated to flip the membrane. This field is on for a period of $\pi/\omega_s$ to flip the membrane orientation, where $\omega_s > \omega$. Once the membrane resembles the starting configuration, the 2-step field is repeated. After half the orbit from Fig. 5b is obtained, the membrane's center of mass has shifted $\sim 2 \rho$. The ``flip" from the second field places the membrane back into its original configuration. This recovery stroke moves the membrane back towards its original position, but not entirely, leading to a net translation. The chirality and duration of the magnetic field precession controls the displacement in the membrane plane and the flip direction controls the direction for the out-of-plane displacement. The fastest achievable velocity using this method is $V_{max}=(2/\pi)V$, but will be slowed by the time taken during the recovery step. 

This cycle forms a closed loop in configuration space based on the two independent degrees of freedom that are defined by the angles the normal vector $\bm{n}$ and the truncation vector $\bm{S}$ make with the x-axis as $\zeta_n$ and $\zeta_S$, respectively (Fig. 6b). It is reasonable to note that this configuration loop is in addition to the already present non-reciprocal motion of the wobbling mode, but is needed since the latter only follows circular paths. Thermalizing the LB fluid to 1 $k_BT$, by the method of Adhikari et al. \cite{Adhikari2005fluctuating} for $S^{3/2}/(\tau \omega)^2 \approx$ 10$^{-2}$, shows a swimming membrane as $\zeta_n$ and $\zeta_S$ changes (Fig. 6c). In this instance, the path during the rotation step, to change $\zeta_S$, is dominated by Brownian motion. The largest displacement occur during the flipping step, to change $\zeta_n$. Additionally, each flip shifts the membrane along the $z$-axis, where the traveling direction is determined by the handedness of the flip. By controlling the precession axis orientation, a membrane may be directed along an arbitrary path.

The useful swimming regime is bound by the P\'{e}clet number and the dimensionless transition frequency. In other words, the system parameters, in particular, the field frequency $\omega$, must be large enough to maintain the wobbling mode, but not too large as to attenuate the wobble amplitude below an efficient swimming velocity. In practice, this implies operating at a driving frequency just above $\omega_c$. The range for the frequency can be written as $2\Omega/\pi<\omega<C^2\eta R^3S^{3/2}/\sqrt2 ~\zeta(3)\tau^2k_BT$. Here, we calculate the P\'{e}clet number using swimming velocity from Appendix E, the membrane radius as the characteristic length, and set the diffusion coefficient using the radius of gyration of a disk \cite{capuani2006disks}. For example, a membrane of $R=$ 1 $\upmu$m composed of 25 nm magnetite nanoparticles at $25^{\circ}$C in water subject to 50 mT field \cite{susan2019from} precessing at 80$^\circ$ gives a effective frequency range of 1--10 kHz.

\section{SWIMMING IN HOMOGENEOUS MEMBRANES}

Homogeneous superparamagnetic membranes require both non-reciprocal motion and shape asymmetry to swim in viscous fluids. While the Scallop Theorem \cite{purcell1977life} makes the necessity for non-reciprocal motion known, implementing such motion without modifying the elastic or magnetic homogeneity implies using a ``non-reciprocal" magnetic field, where the field vector returns to its starting position without retracing its path. Using a field that does not self-retrace imparts a change in membrane conformation that breaks time-reversal symmetry. However, this type of external magnetic field will still generate centrosymmetric forces within a symmetric membrane. Therefore, shape asymmetry is also needed to displace the membrane center of mass during one precession period, where more asymmetry leads to a larger per-period displacement.

\acknowledgements{We would like to acknowledge Mykola Tasinkevych and Eleftherios Kyrkinis for helpful discussions. We thank the Sherman Fairchild Foundation for computational support. This project was funded by the Department of Energy's Center for Bio-Inspired Energy Science (DE-SC0000989).}


\appendix

\section{MD simulations}
The membrane is composed of a hexagonally, close-packed monolayer of hard spherical colloids, of mass $m$ and diameter $\sigma$, enclosed within a circle of radius $R$. The radius is reported in units of $\sigma$. 
Each colloid is connected to its nearest neighbors by inextensible bonds. The bending of the membrane is determined by the quadratic dihedral potentials between each colloid and its neighboring particles
\begin{equation}
U_{bend} =  \frac{\kappa}{2} \sum^N_i \sum^n_{j\in P}{\varphi_{ij}^2}
\end{equation}
where $\kappa$ is the bending rigidity, and $\varphi_{ij}$ is the dihedral angle formed from each colloid $i$ and $j$th unique set of 3 nearest neighbors in $P$, for the number of elements in $P$, $n$. Integrating over all colloids, $N$, yields the total bending energy. The energy scale for $\kappa$ is given by the energy unit $\epsilon$.

A magnetic point dipole of magnitude $\mu$ is placed at the center of each colloid. The precessing magnetic field determines the dipole orientation and is set with a right-hand rotation as
\begin{equation}
\bm{\hat{\mu}} =  \sin{\theta} \sin{\omega t}\, \textbf{\^{i}} +  \sin{\theta} \cos{\omega t}\, \textbf{\^{j}} +  \cos{\theta} \, \textbf{\^{k}}.
\label{eq:mu}
\end{equation}
where $\theta$ is the field precession angle, $\omega$ is the precession frequency, and $t$ is time. The timescale of the simulation is given in units of $t^\ast = \sigma \sqrt{m/\epsilon}$, with the simulation timestep $\Delta t = 10^{-3}~ t^\ast$. 

The potential energy contribution to the total energy is the sum of the dipole-dipole potential energy $U_{dipole}$ over all colloids. The dipole-dipole interaction is given by
\begin{equation}
U_{dipole} = \sum^N_{i} \sum^{N^{\prime}}_{j} \frac{\mu_0 \mu^2}{4 \pi r_{ij}^3} \bigg(1 - 3(\bm{\hat{\mu}} \cdot \bm{\hat{r}_{ij}})^2\bigg)    
\end{equation}
where $\mu$ is the magnitude of the magnetic moment and $\bm{\hat{r}_{ij}}$ is the displacement vector between colloids $i$ and $j$. The length $\bm{\hat{r}_{ij}}$ is given in units of $\sigma$, and $\mu$ is reported in units of $\sqrt{\mu_0/4 \pi \sigma^3 \epsilon}$, where $\mu_0$ is the magnetic permeability. The dipole-dipole interaction is cutoff for all colloids $r_{ij} > 10$. Finally, the motion of the colloids are damped by a drag force, $-\xi v(t)$, proportional to the colloid velocity, $v(t)$, and the damping coefficient $\xi=5 \times 10^{2}~ m/t^\ast$. 

The dynamic states of the membrane were determined by calculating the total potential energy of the membrane over $10^2$ precession periods. The transition frequency was determined to within an error of $\pm 10^{-3}$. The transition was confirmed by taking the Fourier transform of the beads position data over time, $\mathcal{F}(x(t))$, $\mathcal{F}(y(t))$, $\mathcal{F}(z(t))$. 

The MD simulations were performed using LAMMPS \cite{plimpton1995} and snapshots of the membrane were visualized using the software OVITO \cite{ovito}.
\section{Finding the critical frequency transition for a rotating membrane}

We derive the dimensionless transition frequency $\omega_c/\Omega$ from Figure 2. A rigid disk of radius $R$ and thickness $\sigma$ moves in response to a magnetic field $\bm{\hat{H}}=\bm{\hat{\mu}}$. The motion of the disk is synchronous to the field and is visually similar to the rotational wobble of an Euler disk. The disk orientation is defined by the central normal vector on the face of the disk $\bm{\hat{n}} =  -\sin{\gamma} \cos{\omega t}\, \textbf{\^{i}} +  \sin{\gamma} \sin{\omega t}\, \textbf{\^{j}} +  \cos{\gamma} \, \textbf{\^{k}}$ where $\gamma$ is the angle of precession around the $z$-axis for $\bm{\hat{n}}$. We can solve for $\gamma$ in the steady-state by solving for the motion of an arbitrary vector in the membrane plane using the Euler-Lagrange (EL) equation with a velocity-dependent (Rayleigh) dissipation function \cite{goldstein}. Here, we define a second angle $\alpha(t)$, where $\alpha(0)=\gamma$ implying that the vector defined by this angle points along the $x$-direction (this turns out to be the nearest neighbors in the $x$-direction $\pm \bm{\hat{r}_{x}}$). The EL equation to solve for the dynamics of the disk using the degree of freedom $\alpha(t)$ is written as
\begin{equation}
    \frac{d}{dt} \frac{\partial L}{\partial \dot{\alpha}} -\frac{\partial L}{\partial \alpha} + \frac{\partial P}{\partial \dot{\alpha}} = 0
    \label{eq:EL}
\end{equation}
where $\alpha = \alpha(t)$, $\dot{\alpha}=d\alpha/dt$, $P$ is the dissipation function, and $L$ is the Lagrangian. The Lagrangian is defined as the difference $T-U$, where $T$ is the kinetic energy, and $U$ is the potential energy. Now we can expand  Eq. \ref{eq:EL} as
\begin{equation}
    \frac{d}{dt} \frac{\partial T}{\partial \dot{\alpha}} -\frac{d}{dt} \frac{\partial U}{\partial \dot{\alpha}} -\frac{\partial T}{\partial \alpha} + \frac{\partial U}{\partial \alpha} + \frac{\partial P}{\partial \dot{\alpha}} = 0
\end{equation}
Since $\frac{\partial T}{\partial \alpha} = 0$ and angular acceleration is negligible in the Stokes limit,($ \frac{d}{dt} \frac{\partial T}{\partial \dot{\alpha}} \approx 0$ and $ \frac{d}{dt} \frac{\partial U}{\partial \dot{\alpha}} \approx 0$), we are left with 
\begin{equation}
    \frac{\partial U}{\partial \alpha} + \frac{\partial P}{\partial \dot{\alpha}} = 0
\label{eq:ELreduced}
\end{equation}

We take $P$ as the Rayleigh dissipation function defined as
\begin{equation}
    P = \frac{1}{2} \sum^N_i \xi (y \dot{\alpha})^2
\end{equation}
where the sum is over all $N$ colloids $i$ in the disk, $y$ is the distance from the rotation axis and $\xi = 3 \pi \eta \sigma$, where $\eta$ is the dynamic viscosity and $\sigma$ is the colloid diameter. The sum over all colloids is given by the integral
\begin{equation}
   P = \frac{2 \xi}{\sigma^2} \int^R_0 \int^{\sqrt{R^2 -x^2}}_0 (y \dot{\alpha})^2 dy dx= \frac{3 \pi^2 \eta R^4}{4\sigma} \dot{\alpha}^2
\end{equation}

To calculate the total potential energy $U$, we only need to consider the contribution by the magnetic potential energy. For simplicity, we use a nearest neighbor approximation on a hexagonal lattice and sum the magnetic interactions over all colloids in the membrane.
\begin{equation}
   U = \frac{\mu_0}{4\pi}\frac{\mu^2}{\sigma^3} \sum^N_i \sum^n_j (1-3(\bm{\hat{\mu}} \cdot \bm{\hat{r}})^2)
   \label{eq:V}
\end{equation}
where the sum is over each colloid $i$ in the disk of $N$ colloids, for all $n$ nearest neighbors indexed by $j$, $\mu_0$ is the magnetic permeability of free space, $\mu$ is the dipole magnitude, and $\sigma$ is the nearest neighbor distance. Each colloid on a hexagonal lattice has a set displacement vectors, $\bm{\hat{r}}$, that point towards its 6 nearest neighbors: 
$\bm{\hat{r}_{x}} = \pm (\sqrt{1-C_x^2}, 0, C_x)$, 
$\bm{\hat{r}_{y1}} = \pm (\frac{\sqrt{1-C_1^2}}{2},   \frac{\sqrt{3-3C_1^2}}{2}, C_1)$,
$\bm{\hat{r}_{y2}} = \pm (\frac{\sqrt{1-C_2^2}}{2}, - \frac{\sqrt{3-3C_2^2}}{2}, C_2)$, 
where 
$C_x = -\tan{\alpha}$,
$C_1 = -(\frac{1}{2} + \frac{\sqrt{3}}{2}\tan{\omega t})\tan{\alpha}$,
$C_2 = -(\frac{1}{2} - \frac{\sqrt{3}}{2}\tan{\omega t})\tan{\alpha}$. We ignore a correction for the colloids along the perimeter by assuming $2 \pi R / \pi R^2$ is small for large $R$. Plugging in $\bm{\hat{\mu}}$ and the displacement vectors $\bm{\hat{r}_i}$, into Eq. \ref{eq:V}, we expand $U$ in a Taylor series to the first order about the point $\alpha = 0$ (the small wave amplitude regime) to yield
\begin{equation}
   U = A_0 + \frac{9  R^2 \mu_0 \mu^2 \sin{2 \theta} \sin{\omega t}}{2 \sigma^5}  \tan{\alpha} + \mathcal{O}(\alpha^2)
\end{equation}

The lower order terms in $A_0$ are not dependent on $\alpha$ and therefore can be ignored. Additionally, the accuracy of $U$ can be increased by summing the contribution along the same direction as the nearest neighbors. While explicitly adding the contribution to $U$ by including the dipole-dipole interactions between second nearest neighbors, third and so on. Doing so rapid increases the complexity of $U$. Since the contribution to the potential energy decreases as $1/r^3$ on a regular lattice $\frac{1}{r^3}(1+\frac{1}{2^3}+\frac{1}{3^3}+...)$, the sum can be modified by the Riemann zeta function, $\zeta(3) \approx  1.202$. Therefore, to better approximate the full potential, we replace $U$ with $U' = \zeta(3)U$.

Plugging $U'$ and $P$ into Eq. \ref{eq:ELreduced}, we obtain
\begin{equation}
   \dot{\alpha} + \Omega_l \sec^2{\alpha} \sin{\omega t} = 0
\end{equation}
where $\Omega_l = 6 \zeta(3) \mu_0 \mu^2 \sin{2\theta}/\pi^2 \sigma^4 \eta R^2$ for a system with simple friction of each colloid. For a hydrodynamic system, the potential dissipation function $P = \frac{1}{2} \xi \dot{\beta}^2$ is taken to be consistent with the torque acting on a rotating disk in a viscous fluid, $\xi = \frac{32}{3} \eta R^3$. This changes the size scaling dependence due to hydrodynamic interactions to yield $\Omega_h = 27 \zeta(3) \mu_0 \mu^2 \sin{2\theta}/64 \sigma^5 \eta R$. The inverse of $\Omega_h$ represents the timescale of the membrane's magnetoviscous response $\tau = 1/\Omega_h$ and is part of the magnetoviscous parameter $\tau \omega$. Solving this equation for $\alpha$,
\begin{equation}
 \frac{1}{2}(\alpha + \frac{\sin{2 \alpha}}{2}) = \frac{\Omega_i}{\omega}\cos{\omega t} + Z
    \label{eq:LG2}
\end{equation}
where $i=l$ or $h$, and $Z=0$ with the boundary condition $\alpha=0$ at $\omega t = -\pi/2$. The dynamic transition occurs at $\alpha \rightarrow \pi/2$ during the maximum possible amplitude over a field precession. During steady-state conditions this occurs at $\omega t = 0$. Doing so leads to a simple relation for the transition frequency
\begin{equation}
  \frac{\omega_c}{\Omega} = \frac{2}{\pi}
\end{equation}

\section{Lattice Boltzmann Method coupling to MD model}
Hydrodynamic interactions around the membrane is modeled using the lattice Boltzmann method (LBM) \cite{Mackay2013hydrodynamic}. LBM calculates the evolution of a discrete-velocity distribution function, $f_i$, at each fluid node that fills the simulation box on a square lattice mesh with a spacing of $\Delta x$. Using the Bhatnagar-Gross-Krook (BGK) collision operator \cite{BGK1954model} lattice Boltzmann equation reads
\begin{equation}
\begin{split}
f_i(\bm{x} + \bm{c}_i\Delta t, t + \Delta t) &- f_i(\bm{x}, t) \\
&= -\frac{\Delta t}{\tau}(f_i(\bm{x}, t)-f_i^{eq}(\bm{x}, t)) + W_i
\end{split}
\end{equation}
The left hand side describes the fluid streaming from one node to neighboring nodes along the velocity $\bm{c}_i$. The first term on the right hand side describes the relaxation of the distribution $f_i$ to the equilibrium distribution $f_i^{eq}$, where $\tau$ is the relaxation time. The $W_i$ term defines external forces on the fluid. These forces are distributed from nodes that define the surface of the colloids, $\sigma/2$ distance away from the particle center. These nodes interact with the fluid using the immersed boundary model developed by Peskin \cite{peskin2002immersed}. 

The fluid parameters are set to reproduce the frictional coefficient of a spherical colloid, which accurately reproduces the drag on a disk. The disk is immersed within a box of $112 \times 112 \times 113$ nodes. The spacing between nodes was set as $\Delta x = \sigma/2$ while the time step was set to match the MD simulations. Decreasing Reynolds numbers in LBM can be achieved by choosing a increasing the lattice spacing $\Delta x$, increasing the time relaxation parameter or decreasing the Mach number. It is necessary to maintain resolution of the fluid circulation along the membrane edge and avoid numerical errors associated with increasing the relaxation parameter. Therefore, we rely on a small timestep $\Delta t$ that also integrates effectively with the harmonic potentials in the MD scheme to maintain $Re \ll 1$.

The ``two-step" magnetic field was employed using a truncated membrane ($S=0.48$). The temperature of the LB fluid was thermalized to $1~k_BT$. \cite{Adhikari2005fluctuating} The first step ran for 21 precession periods with $S^{3/2}/(\tau \omega)^2 = 2 \times 10^{-2}$. The second step used set precession angle to $\pi/2$ and rotated the precession axis at a frequency $\omega / \omega_s \approx 10$ for roughly 1/3 the duration of the first step. Because the wobbling mode is stable, the flip does not need to lie perfectly in the x-y plane for the process to repeat effectively.

The parameters for estimating the swimming speed in the main text are taken as typical values for a superparamagnetic system in water. The fundamental constants are set as: $\sigma = 25$ nm, $m = 4.2 \times 10^{-20}$ kg, and $\epsilon$ = k$_B$T at $25^{\circ}$C which corresponds to the kinematic viscosity of water, $\nu =$ 8.9 $\times$ 10$^{-7}$ m$^2$/s. To save computational resources, $\sigma$ was increased to $50$ nm and the magnitude of the dipole and the frequency were increased to 76 Am$^2$/kg and 30 kHz, respectively. We calculate the P\'{e}clet number using swimming velocity from Appendix E, the membrane radius as the characteristic length, and set the diffusion coefficient using the radius of gyration of a disk \cite{capuani2006disks}.

\section{Derivation of the magnetoviscous parameter}

Here, we consider a magnetoelastic membrane in the small amplitude limit. Under small distortions, the bending and stretching forces can be neglected. However, the flexibility of the membrane means that only a small area $\delta a$ near the perimeter is in motion. Therefore, we consider Eq. \ref{eq:ELreduced} over $\delta a$ moving freely along the $z$-direction. We write the drag as simply $\frac{\partial P}{\partial \dot{z}}=3 \pi \eta \sigma  \dot{z} \delta a$. Solving for the resulting first order differential equation yields
\begin{equation}
   \frac{z}{R} = \frac{C\cos{\omega t}}{\tau \omega} 
\end{equation}
where the equation is put in terms of $\tau = 1/\Omega$, and $C = 32/9\pi^2$. The amplitude is maximized $z_{max}=A$ when $\cos{\omega t} = 1$, From this equation, we see that the reduced amplitude of the rotational waves is inversely proportional to the magnetoviscous parameter.

\section{Stokeslet advection of the center of mass}

We use a singularity method to determine the advection of the membrane in a viscous fluid due to point forces generated by magnetic dipole interactions. The fluid velocity at $\bm{x}$ is calculated by summing over all point disturbances \cite{guazzelli_morris_pic_2011}
\begin{equation}
   \bm{v}(\bm{x}) = \frac{1}{8 \pi \eta} \int \bm{G}(\bm{x}-\bm{y}) \cdot \bm{f}(\bm{y}) d\bm{y} 
   \label{eq:uI}
\end{equation}
where $\eta$ is the dynamic viscosity,  $\bm{f}(\bm{y})$ is a point force located at $\bm{y}$, $\bm{G}(\bm{r})=\frac{\bm{I}}{|\bm{r}|} + \frac{\bm{r}\bm{r}}{|\bm{r}|^3}$ is the Oseen tensor, and $\bm{I}$ is the identity matrix.

At every point on the membrane surface, the no-slip fluid velocity is related to the rigid translation $\bm{V}$ and angular rotation $\bm{W}$ of the disk by 
\begin{equation}
    \bm{v}(\bm{x}) = \bm{V} + \bm{W} (\bm{x} - \bm{x}_{com})
    \label{eq:uU}
\end{equation}
where $\bm{x}_{com}$ is the initial center of mass of the disk located at the origin. Since we are interested in the center of mass advection, we can ignore the rotation $\bm{W}$ of the membrane and solve Eq. \ref{eq:uI} to obtain $\bm{V}$.

Equation \ref{eq:uI} can be simplified by considering the underlying lattice and membrane symmetry. For colloids arranged in a lattice along a rigid disk, the magnetic dipole force cancels out for all colloids in the bulk when considering only nearest neighbors. Therefore, the forces along the perimeter dominate. Furthermore, the center of inversion symmetry for all points along the perimeter of a circular membrane yields opposite forces of equal distance from the center resulting in a stationary membrane, $\bm{v}(\bm{x}_{com})=0$. Therefore, we create an imbalance of point forces by making a small truncation made along the $y$-direction in the $x>0$ domain, (i.e. we remove a circular segment centered along the $x$-axis). The difference in the number of point forces due to truncation $\Delta N$ on opposite sides of the membrane is due to the number of colloids that can fit along the difference in the edge lengths $\Delta L/\sigma$, where $\Delta L = 2R[2\sqrt{S(1-S)} - \sin^{-1}({2\sqrt{S(1-S)}})]$. By using the Puiseux series, we can  approximate $\Delta N = 8 R S^{3/2} / 3 \sigma$, which is accurate for $S < (3/8)^{2/3} \approx 0.52$. Since the truncation is small, the point force imbalance occurs along the vector $\bm{x} = - R \bm{\hat{r}_{x}}$. We calculate $\bm{V}$ = $\bm{v}(\bm{x}_{com})$ for an infinitesimally thin disk. With these simplifications, Eq. \ref{eq:uI} and takes the form
\begin{equation}
\bm{V}  = \frac{\Delta N (\bm{\hat{\mu}} ~\cdot ~\bm{\hat{r}_{x}})}{8 \pi \eta R} (\bm{I} + \bm{\hat{r}_{x}}\bm{\hat{r}_{x}}) \cdot \bm{\hat{\mu}}
\end{equation}

Time-averaging $\bm{v}$ over the precession period $2 \pi/\omega$, we obtain
\begin{equation}
<V_x>~=~<V_z>~=~0
\end{equation}

\begin{equation}
<V_y> =  \frac{S^{3/2}\mu_0\mu^2 \sin{2\theta} \tan{\gamma}}{8\pi^2 \eta \sigma^5} 
\end{equation}
We can take replace reduced amplitude $\tan \gamma$ with the results from Appendix D to obtain a nondimensional swimming velocity reduced by $R \omega$.
\begin{equation}
\frac{<V_y>}{R \omega} = 
 \frac{C^2}{12 \zeta(3)}\frac{S^{3/2}}{(\tau \omega)^2} 
\end{equation}

For the two-step linear swimmer, the theoretical maximum swimming speed, assuming no distance loss during the recovery stroke is 
\begin{equation}
    V_{max} = \frac{2}{\pi} <V_y>
\end{equation}

\bibliography{biblio}

\end{document}